\documentclass[iop]{emulateapj}
\usepackage{amsmath}
\usepackage{epsfig}
\usepackage{natbib}
\usepackage{color}
\usepackage{multirow}

\newcommand{\beq}{\begin{equation}}
\newcommand{\eeq}{\end{equation}}

\shorttitle{Astrophysical factors for Ret~II}
\shortauthors{Bonnivard et al.}

\def\nar{New Astronomy Review}%

\begin{document}

\title{Dark matter annihilation and decay profiles for the Reticulum II dwarf spheroidal galaxy}

\author{Vincent~Bonnivard\altaffilmark{1}}
\email{bonnivard@lpsc.in2p3.fr}
\author{C\'eline~Combet\altaffilmark{1}}
\author{David~Maurin\altaffilmark{1}}
\author{Alex Geringer-Sameth\altaffilmark{2}}
\author{Savvas M. Koushiappas\altaffilmark{3}}
\author{Matthew G. Walker\altaffilmark{2}}
\email{mgwalker@andrew.cmu.edu}
\author{Mario Mateo\altaffilmark{4}}
\author{Edward W. Olszewski\altaffilmark{5}}
\author{John I. Bailey III\altaffilmark{4}}
\altaffiltext{1}{LPSC, Universit\'e Grenoble-Alpes, CNRS/IN2P3, 53 avenue des Martyrs, 38026 Grenoble, France}
\altaffiltext{2}{McWilliams Center for Cosmology, Department of Physics, Carnegie Mellon University, Pittsburgh, PA 15213, USA}
\altaffiltext{3}{Department of Physics, Brown University, Providence, RI 02912, USA}
\altaffiltext{4}{University of Michigan, 311 West Hall, 1085 S. University Ave., Ann Arbor, MI 48109, USA}
\altaffiltext{5}{Steward Observatory, The University of Arizona, 933 N. Cherry Ave., Tucson, AZ 85721, USA}

\begin{abstract}
The dwarf spheroidal galaxies (dSph) of the Milky Way are among the
most attractive targets for indirect searches of dark matter. In this
work, we reconstruct the dark matter annihilation ($J$-factor) and
decay profiles for the newly discovered dSph Reticulum~II. Using an
optimized spherical Jeans analysis of kinematic data obtained from the
Michigan/Magellan Fiber System (M2FS), we find Reticulum~II's $J$-factor to be among the largest of any Milky Way dSph.
We have checked the robustness of this result against several ingredients of the analysis. Unless it suffers from tidal disruption or significant inflation of its velocity dispersion from binary stars, Reticulum~II may provide a unique window on dark matter particle properties. 
\end{abstract}

\keywords{galaxies: dwarf --- galaxies: individual (Reticulum~II) --- dark matter --- gamma rays: galaxies --- methods: statistical --- stars: kinematics and dynamics}

\section{Introduction}

Along with the Galactic center and galaxy clusters, the dwarf spheroidal galaxies (dSph) of the Milky Way have been identified as promising targets for indirect dark matter (DM) searches (see e.g. \citealt{2013PhR...531....1S,2015arXiv150306348C}). Their low astrophysical background, high mass-to-light ratio, and proximity make them compelling targets \citep{1990Natur.346...39L,2004PhRvD..69l3501E}. About twenty-five Galactic dSphs were known as of early 2015, and their observation by $\gamma$-ray telescopes has thus far shown no significant emission, leading to stringent constraints on $\langle\sigma_{\rm ann} v\rangle$, the thermally-averaged DM annihilation cross-section \citep{2010ApJ...720.1174A,2011arXiv1110.6775P,2014PhRvD..90k2012A,2014arXiv1410.2242G,2015arXiv150302641F}.

Recently, imaging data from the Dark Energy Survey has led to the discovery of nine new potential Milky-Way satellites in the Southern sky \citep{2015arXiv150302079K,2015arXiv150302584T}. The nearest object, Reticulum~II (Ret~II, $d \sim 30$ kpc), is particularly intriguing, as evidence of $\gamma$-ray emission has been detected in its direction using the public Fermi-LAT Pass 7 data. \citet{2015arXiv150302320G} determined the probability of background processes producing the observed Ret~II gamma-ray signal to be between $p = 0.01\%$ and $p = 1\%$, depending on the background modelling. An analysis of the new objects published simultaneously by \citet{2015arXiv150302632T}, based on the unreleased Pass 8 data set, reported no significant detection, though the strongest hint was for Ret~II with $p = 6\%$. \citet{2015arXiv150306209H} subsequently performed a similar analysis with public Pass 7 data, finding a $p$ value of $0.16\%$.

In any case, a robust determination of Ret~II's DM content is crucial
in order to constrain particle nature of DM. Reticulum~II was
  found to be a DM-dominated dSph galaxy from the independent
  chemodynamical analyses of \citet{kinematics},
  \citet{2015arXiv150402889S} and \citet{2015arXiv150407916K}. Here,
we reconstruct the DM annihilation and decay emission profiles of
Ret~II from a spherical Jeans analysis applied to stellar kinematic
data obtained with the Michigan/Magellan Fiber System (M2FS)
\citep{kinematics}. We use the optimized Jeans analysis setup from
\citet{2015MNRAS.446.3002B,2015arXiv150402048B}, and compute the
astrophysical $J$- and $D$-factors, for annihilating and decaying DM
respectively, from the reconstructed DM density profiles. We
cross-check our results by varying different ingredients of the
analysis and evaluate the ranking of Ret~II among the most promising dSphs for DM indirect detection.

\section{Astrophysical factors, Jeans analysis and data sets}
\label{sec:jeans}
\subsection{Astrophysical factors}
The differential $\gamma$-ray flux coming from DM annihilation
(resp. decay) in a dSph galaxy is proportional to the so-called 'astrophysical factor' $J$ (resp. $D$) \citep{1998APh.....9..137B},
\begin{equation}
      \!J \!=\! \!\int\!\!\!\!\!\int\!\! \rho_{\rm DM}^2 (l,\Omega)
      \,dld\Omega  \!\!\!\!\quad\left({\rm \!resp.~} D \!=\!\!
      \int\!\!\!\!\!\int \!\!\rho_{\rm DM}(l,\Omega)
      \,dld\Omega\!\!\right)\!,\!\!
      \label{eq:J}
 \end{equation} 
which corresponds to the integration along the line-of-sight (l.o.s.) of the DM
density squared (resp. DM density) and over the solid angle $\Delta\Omega
= 2\pi\times[1-\cos(\alpha_{\rm int})]$, with $\alpha_{\rm int}$ the
integration angle. This quantity depends on both the extent of the DM halo and the mass density distribution, and is essential for constraining the DM particle properties. All calculations of astrophysical factors are done with the {\tt
CLUMPY} code \citep{2012CoPhC.183..656C}, a new module of which has been
specifically developed to perform the Jeans analysis\footnote{This upgrade
will be publicly available in the new version of the software (Bonnivard et al., in prep.).}.

\subsection{Jeans analysis}
\label{subsec:jeans}
Several approaches have been developed to infer the DM density profile
of dSph galaxies from stellar kinematics (see
e.g. \citealt{2013NewAR..57...52B,2013PhR...531....1S,2013pss5.book.1039W}). Here,
we focus on the spherical Jeans analysis, a widely-used approach
for the determination of astrophysical factors
\citep{2007PhRvD..75h3526S,2010PhRvD..82l3503E,2011MNRAS.418.1526C,2012PhRvD..86b3528C,2015ApJ...801...74G,2015arXiv150402048B}. We
refer the reader to \citet{2015MNRAS.446.3002B} for a thorough
description of the analysis setup we use in this work. Here, we summarize the main ingredients.

Assuming steady-state, spherical symmetry, and negligible rotational support, the second-order Jeans equation, obtained from the collisionless Boltzmann equation, reads \citep{2008gady.book.....B}:
\begin{equation}
\frac{1}{\nu}\frac{d}{dr}(\nu \bar{v_r^2})+2\frac{\beta_{\rm ani}(r)\bar{v_r^2}}{r}=-\frac{GM(r)}{r^2},
\label{eq:jeans}
\end{equation}
with $\nu(r)$ the stellar number density, $\bar{v_r^2}(r)$ the radial velocity dispersion, $\beta_{\rm ani}(r)\equiv 1-\bar{v_{\theta}^2}/\bar{v_r^2}$ the velocity anisotropy, and $M(r)$ the mass\footnote{The mass is dominated by DM, and we neglect the stellar component.} enclosed within radius $r$. After solving Eq.~(\ref{eq:jeans}) and projecting along the l.o.s., the (squared) velocity dispersion at the projected radius $R$ reads
\begin{equation}
  \sigma_p^2(R)=\frac{2}{\Sigma(R)}\displaystyle \int_{R}^{\infty}\biggl (1-\beta_{\rm ani}(r)\frac{R^2}{r^2}\biggr ) 
  \frac{\nu(r)\, \bar{v_r^2}(r)\,r}{\sqrt{r^2-R^2}}\mathrm{d}r,
  \label{eq:jeansproject}
\end{equation}
with $\Sigma(R)$ the surface brightness profile. We compare the l.o.s velocities of the stars to the projected velocity dispersion $\sigma_p$, computed using parametric forms for the unknown velocity anisotropy $\beta_{\rm ani}(r)$ and DM density profile $\rho_{\rm DM}(r)$. We use the following likelihood function \citep{2007PhRvD..75h3526S}
\begin{equation}
  \mathcal{L}\!=\! \!\prod_{i=1}^{\!\!N_{\rm stars}}
  \! \!\!\frac{(2\pi)^{-1/2}}{\sqrt{\sigma_p^2(R_i)\!+\!\Delta_{v_{ i}}^{2}}}
  \exp\!\biggl [\!-\frac{1}{2}\biggl (\!\frac{(v_{\rm i} \!-\!\bar{v})^{2}}{\sigma_p^2(R_i)\!+\!\Delta_{v_{i}}^{2}\!}
  \biggr )\biggr ] \!,
  \label{eq:likelihood}
\end{equation}
which assumes a Gaussian distribution of l.o.s. stellar velocities $v_i$, centered on the mean stellar velocity $\bar{v}$, with a dispersion of velocities (at the radius $R_i$) coming from both the intrinsic dispersion $\sigma_p(R_i)$ and the measurement uncertainty $\Delta_{v_{i}}$. Probability density functions (PDFs) of the anisotropy and DM parameters are obtained with a Markov Chain Monte Carlo (MCMC) engine\footnote{We use the {\tt GreAT} toolkit \citep{2011ICRC....6..260P,Putze:2014aba}.}, and are used to compute the median and credible intervals (CIs) of the astrophysical factors for any integration angle.  

Following the {\it optimized} Jeans analysis setup proposed in \citet{2015MNRAS.446.3002B}, the DM density is described by an Einasto profile (\citealt{2006AJ....132.2685M}), and the anisotropy and light profiles are given by  Baes \& van Hese \citep{2007AA...471..419B} and Zhao-Hernquist \citep{1990ApJ...356..359H,1996MNRAS.278..488Z} parametrisations, respectively. The large freedom allowed by these parametrisations was found to mitigate possible biases of the Jeans analysis \citep{2015MNRAS.446.3002B}. Finally, the extent of the DM halo is computed using the tidal radius estimation as in \citet{2015arXiv150402048B}.
\begin{figure}[t]
\begin{center}
\includegraphics[width=\linewidth]{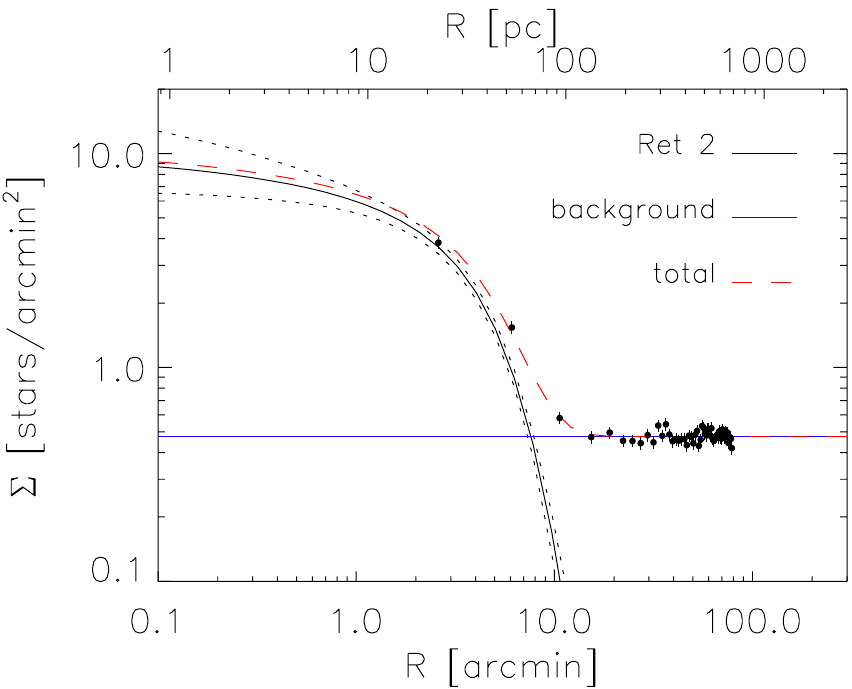}
\caption{Projected stellar density profile of Ret~II, derived from the photometric catalog of \citet{2015arXiv150302079K}.  Overplotted (red line) is the best-fitting model (we note that the fit is to the unbinned data), which is the sum of contributions from Ret~II itself and a constant background (see Section \ref{subsec:data}). Dotted lines enclose 68\% CIs for the projection of $\nu(r)$.}
\label{fig:light}
\end{center}
\end{figure}
\subsection{Data set}
\label{subsec:data}
\paragraph{Surface brightness data}
We fit the stellar number density profile $\nu(r)$ of Ret~II following the procedure that \citet{2015arXiv150402048B} use for `ultrafaint' dSphs (see their section 3.1). We consider a flexible Zhao-Hernquist model for the 3D profile,
\begin{equation}
           \nu^{\rm Zhao}(r)=\frac{\nu_s^{\star}}{(r/r_s^{\star})^\gamma
             [1+(r/r_s^{\star})^\alpha]^{(\beta-\gamma)/\alpha}}\;,
           \label{eq:nu_zhao}
\end{equation}
where the five parameters are the normalization $\nu_s^{\star}$, the
scale radius $r_s^{\star}$, the inner power law index $\gamma$, the outer index $\beta$, and the transition parameter $\alpha$. Along with an additional free parameter $\Sigma_{\rm bkd}$ that represents a uniform background density, these parameters then specify a model for the projected stellar density:
\begin{equation}
  \Sigma_{\rm model}(R)\equiv
  2\displaystyle\int_{R}^\infty\frac{\nu(r)r}{\sqrt{r^2-R^2}}dr+\Sigma_{\rm
    bkd}.
  \label{eq:sbpmodel}
\end{equation}
We fit this model to the photometric catalog generated by \citet{2015arXiv150302079K}, which provides positions, colors, and magnitudes of individual stars detected as point sources.  From the raw catalog, we first identify possible members of Ret~II as point sources (selected as sources with Sextractor `spread' parameter $<$ 0.01 in the $g$-band) whose extinction-corrected $g-r$ colors place them within 0.25 dex of the Dartmouth isochrone \citep{2008ApJS..178...89D}, calculated for a stellar population with age 12 Gyr, metallicity $\rm{[Fe/H]}=-2.5$, and distance modulus $m-M=17.4$ \citep{2015arXiv150302079K}.  To the unbinned distribution of projected positions for the $N=12470$ RGB candidates identified within $1.5^\circ$ of Ret~II's center, we fit 2D projections of $\nu(r)$ according to the likelihood function:
\begin{equation}
  \mathcal{L}_2\propto \displaystyle\prod_{i=1}^{N} \Sigma_{\rm model}(R_i).
\end{equation}
As in \citet{2015arXiv150402048B}, the fit is done with the software package MultiNest \citep{2008MNRAS.384..449F,2009MNRAS.398.1601F,2013arXiv1306.2144F}, and we use the samples from the posterior PDFs to propagate the light profile uncertainty into the Jeans analysis. Figure \ref{fig:light} shows the fit to the projected stellar density profile of Ret~II (dashed red line), with the contributions from Ret~II itself and from the constant background (solid black and blue lines respectively).
\begin{figure}[t]
\begin{center}
\includegraphics[width=\columnwidth]{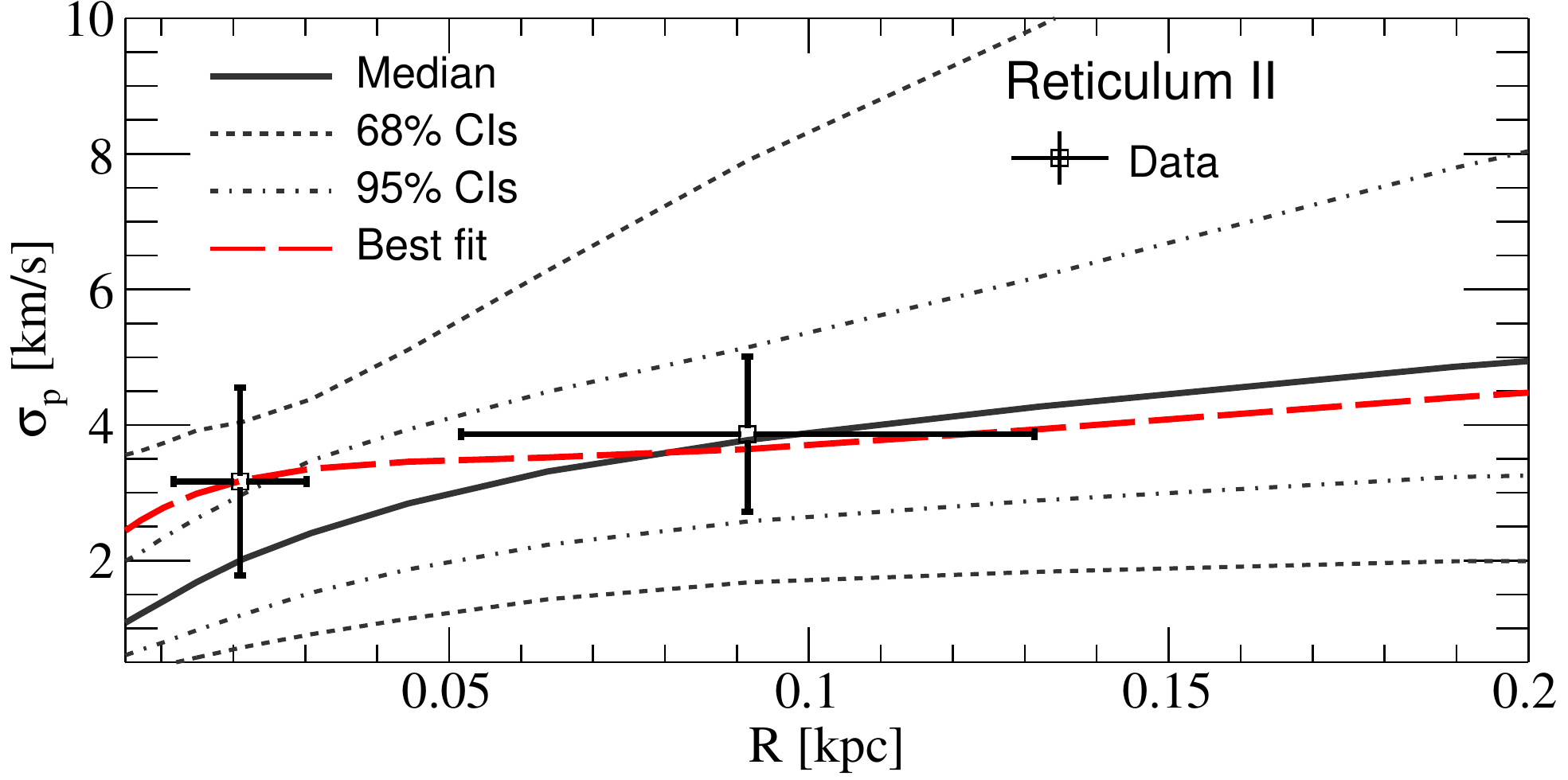}
\includegraphics[width=\columnwidth]{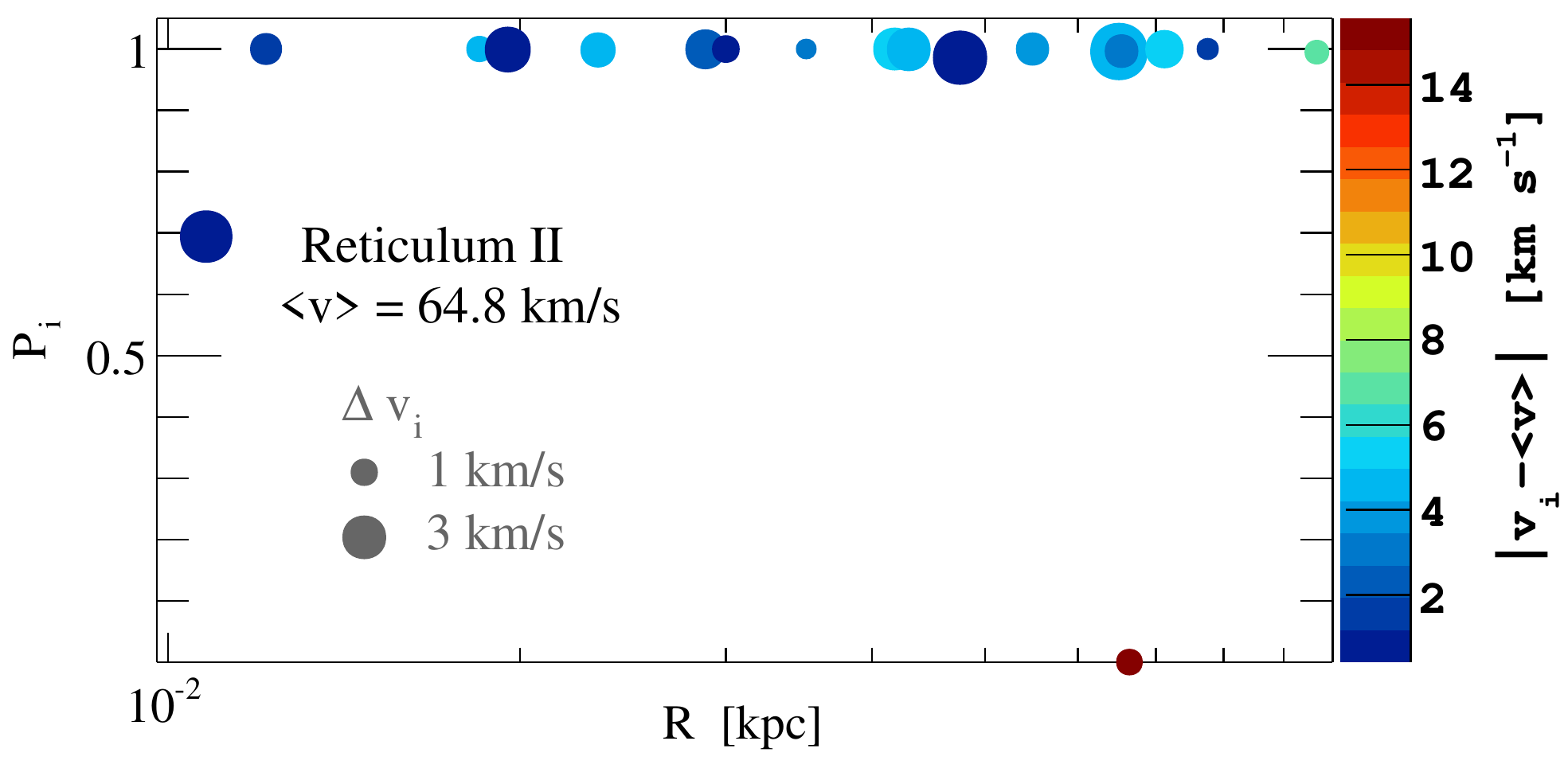}
\caption{{\em Top:} velocity dispersion profile of Ret~II and reconstructed median and credible intervals (solid and dashed black lines respectively), as well as best fit$^9$ (long dashed red lines). {\em Bottom:} distribution of membership probabilities as a function of the projected radius $R$ and the departure from the mean velocity ($z$-axis, blue to red color) for the eighteen stars with $P_i \neq 0$. The size of the points is proportional to the velocity uncertainty. See text for discussion.}
\label{fig:pmem_radius}
\end{center}
\end{figure}
\paragraph{Kinematic data}
We use the Ret~II stellar kinematic data set from \citet{kinematics}, obtained with M2FS. It consists of projected positions and l.o.s. velocities for 38 individual stars, as well as an estimation of their membership probability $P_i$. The latter, obtained using an expectation maximization algorithm \citep{2009AJ....137.3109W}, quantifies the probability that a given star belongs to the dSph or to the Milky Way foreground. 

The top panel of Figure \ref{fig:pmem_radius} presents the velocity dispersion profile of Ret~II, as well as its reconstruction with the Jeans analysis\footnote{The binned data and associated velocity dispersion reconstruction are only shown for illustration purposes. The final results are obtained with an unbinned analysis.}. The bottom panel of Figure \ref{fig:pmem_radius} shows the distribution of membership probabilities as a function of the projected radius $R$ and the departure from the mean velocity (color-coded), for stars with non-zero $P_i$. As pointed out in \citet{2015arXiv150402048B}, a large fraction of stars with both intermediate $P_i$ ($0.1 < P_i < 0.95$) and large departure from the mean velocity hints at Milky Way foreground contamination, which can affect the $J$- and $D$-factor reconstruction. For Ret~II, only one star shows an intermediate $P_i$ (Ret2-142 in the catalog of \citealt{kinematics}, with $P_i = 0.69$), with a very small departure from the mean velocity. Therefore we do not expect a strong sensitivity to foreground contamination. In this study, and as advocated in \citet{2015arXiv150402048B}, we use the data with $P_i > 0.95$ (sixteen likely members, one less than identified by \citealt{kinematics} after exclusion of Ret2-142) as our fiducial setup.
\section{Results}
\label{sec:J}

\begin{table}
\begin{center}
\caption{Astrophysical factors for Ret~II ($d = 30$ kpc). For five different integration angles, the median $J$ (resp $D$)-factors as well as their 68\% and 95\% CIs are given.  Note that possible triaxiality of the dSph galaxies adds a systematic uncertainty  of $\pm0.4$ (resp.~$\pm0.3$) \citep{2015MNRAS.446.3002B} and is not included in the quoted intervals.
\label{tab:results}}
\begin{tabular}{ccccc} \hline \hline
  $\alpha_{\rm int}$ &&   $\log_{10}(J(\alpha_{\rm int}))$  &&  $\log_{10}(D(\alpha_{\rm int}))$ \\[0.1cm]
 [deg] &&$[J/$GeV$^2 \,$cm$^{-5}]$\tablenote{1~GeV$^2 \,$cm$^{-5} = 2.25\times 10^{-7} M_{\odot}^{2}\,$kpc$^{-5}$} &&$[D/$GeV$ \,$cm$^{-2}]$\tablenote{1~GeV$ \,$cm$^{-2} = 8.55\times 10^{-15} M_{\odot}\,$kpc$^{-2}$}\\[0.05cm]
\hline
0.01 && $17.1_{-0.5(-0.9)}^{+0.5(+1.1)}$ && $15.7_{-0.3(-0.5)}^{+0.6(+1.0)}$ \\[0.2cm]
0.05 && $18.3_{-0.4(-0.8)}^{+0.5(+1.1)}$ && $17.0_{-0.3(-0.6)}^{+0.5(+1.0)}$ \\[0.2cm]
0.1  && $18.8_{-0.5(-0.8)}^{+0.6(+1.2)}$ && $17.6_{-0.4(-0.6)}^{+0.6(+1.1)}$ \\[0.2cm]
0.5  && $19.6_{-0.7(-1.3)}^{+1.0(+1.7)}$ && $18.8_{-0.7(-1.1)}^{+0.7(+1.2)}$ \\[0.2cm]
1    && $19.8_{-0.9(-1.4)}^{+1.2(+2.0)}$ && $19.3_{-0.9(-1.4)}^{+0.8(+1.4)}$ \\[0.1cm]
\hline
\end{tabular}
\end{center}
\end{table}

\begin{figure}[t]
\begin{center}
\includegraphics[width=\columnwidth]{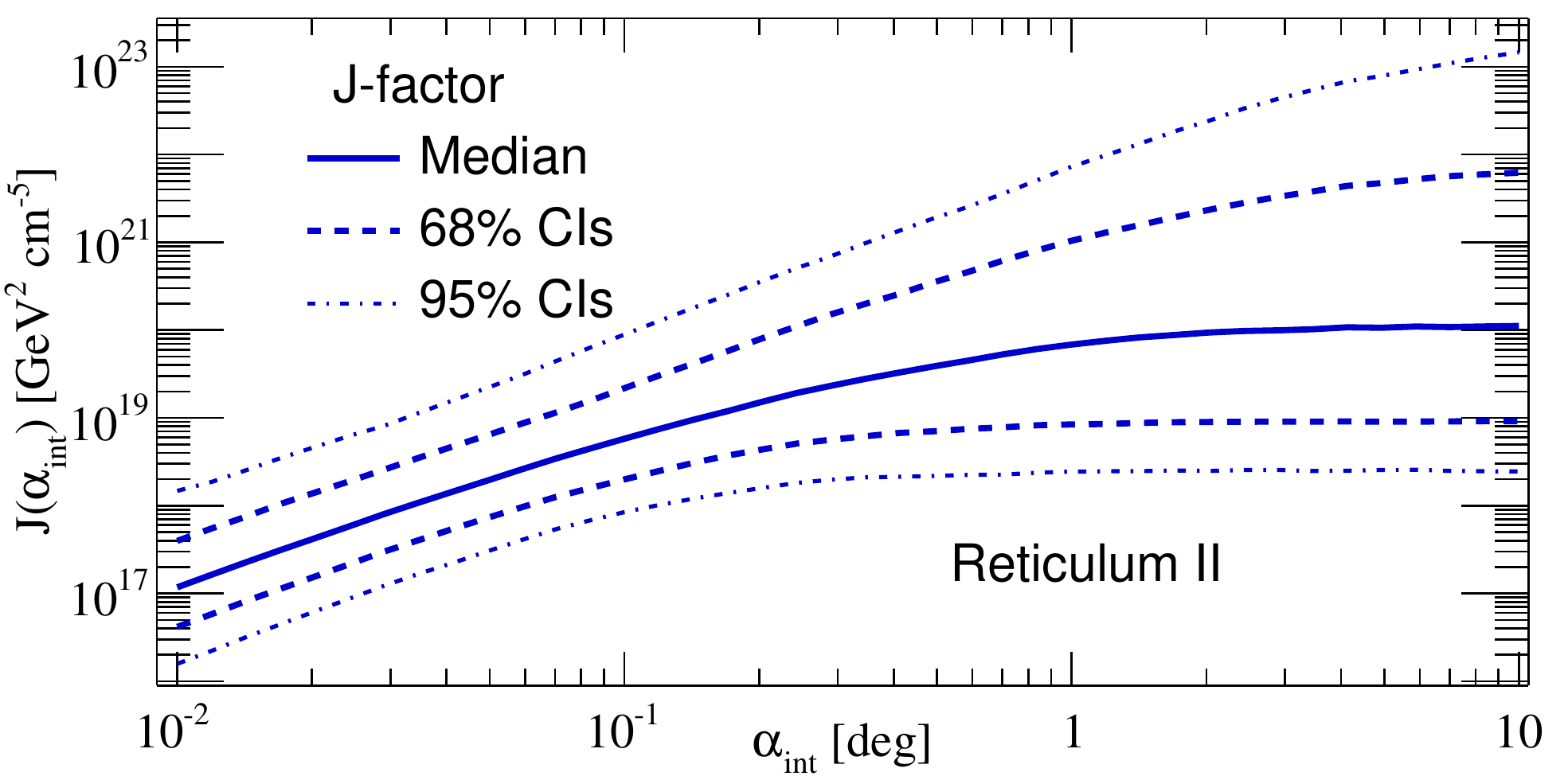}
\includegraphics[width=\columnwidth]{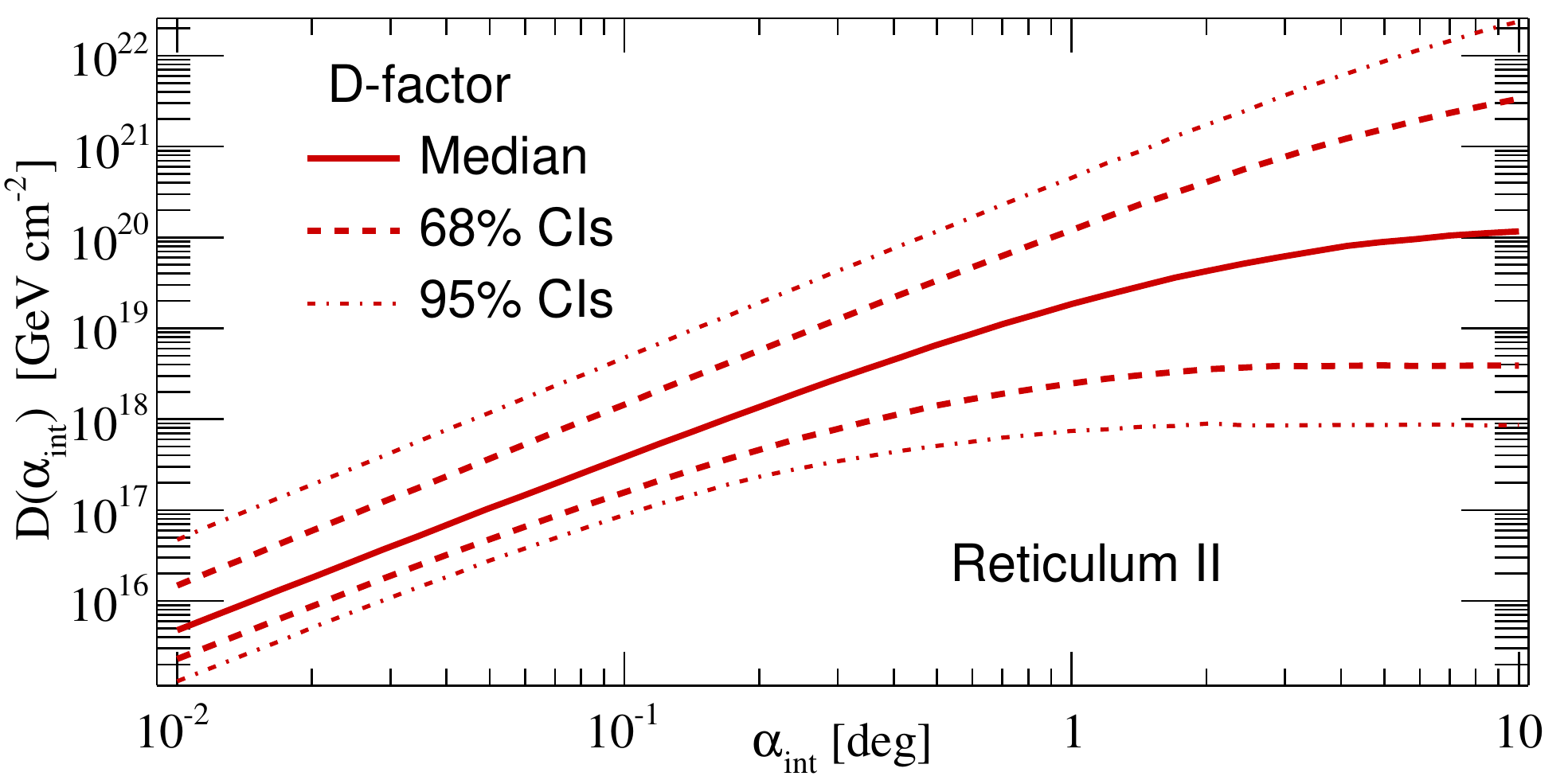}
\caption{Median (solid), 68 \% (dashed), and 95\% (dash-dot) CIs of the $J$- (top) and $D$-factors (bottom) of Ret~II, as a function of integration angle, reconstructed from our Jeans/MCMC analysis.}
\label{fig:J_D}
\end{center}
\end{figure}

\begin{figure}[t]
\begin{center}
\includegraphics[width=\columnwidth]{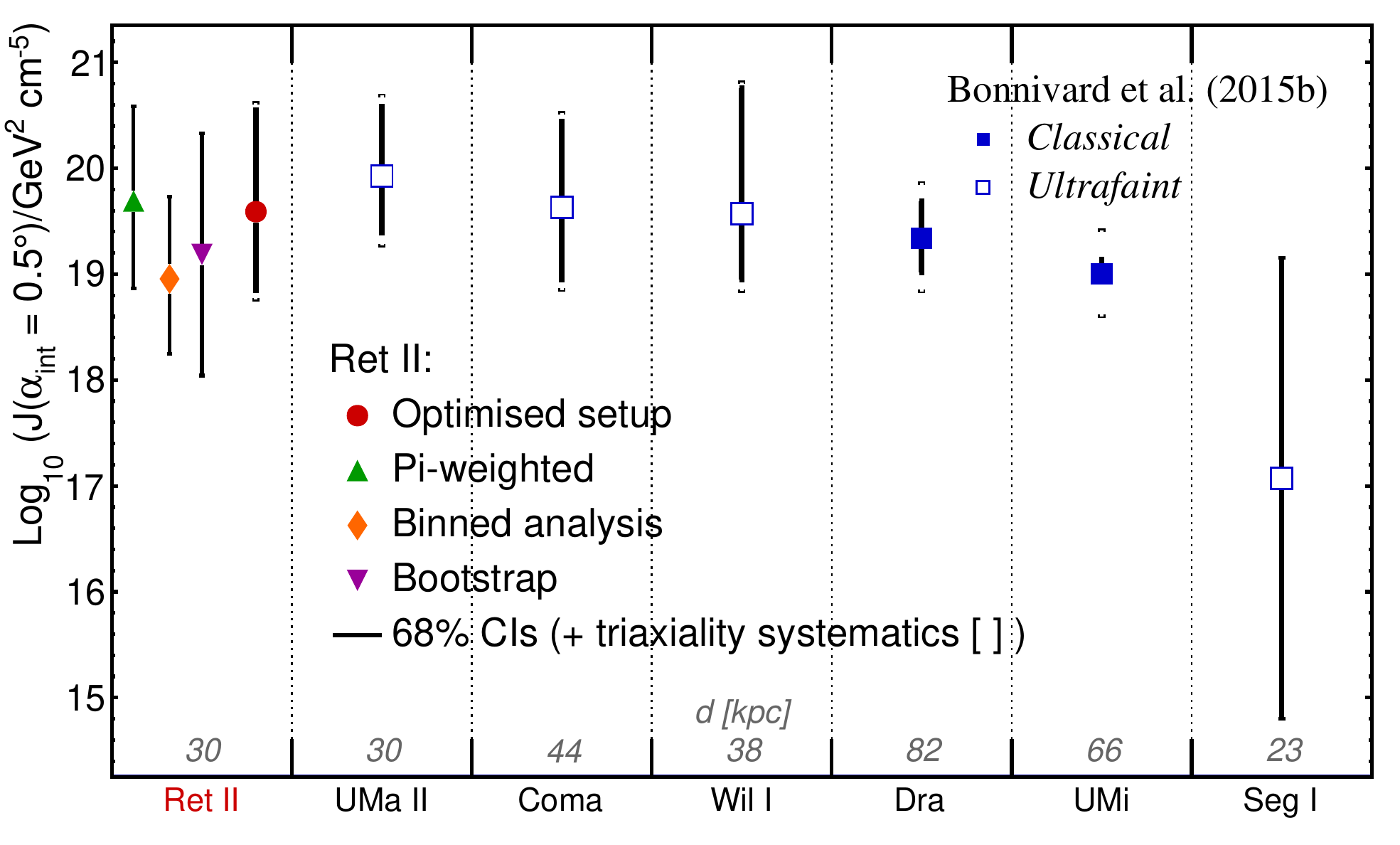}
\caption{Comparison of the $J$-factors at $\alpha_{\rm int} = 0.5^{\circ}$ obtained for Ret~II (red circle) and for the potentially brightest objects from \citet{2015arXiv150402048B} (blue squares), with the same Jeans/MCMC analysis. Ret~II is comparable to Wil~I in terms of $J$-factors, but slightly below Coma and UMa~II. A 0.4 dex systematic uncertainty was added in quadrature to the 68\% CIs to account for possible triaxiality of the DM halo \citep{2015MNRAS.446.3002B}. Also shown are the $J$-factors obtained for Ret~II by varying different ingredients of the analysis - see Section \ref{sec:J}.}
\label{fig:J_Comp}
\end{center}
\end{figure}

Figure \ref{fig:J_D} displays the $J$- (top) and $D$-factors (bottom) of Ret~II, reconstructed from the Jeans/MCMC analysis, as a function of the integration angle $\alpha_{\rm int}$. Solid lines represent the median values, while dashed and dash-dot lines symbolize the 68\% and 95\% CIs respectively. Our data-driven Jeans analysis gives large statistical uncertainties due to the small size of the kinematic sample, comparable to those obtained for other `ultrafaint' dSphs by \citet{2015arXiv150402048B} (see also Figure \ref{fig:J_Comp}). Table \ref{tab:results} summarizes our results for the astrophysical factors of Ret~II.

We cross-check our findings by varying different ingredients of the
Jeans analysis. The resulting $J$-factors are shown in Figure
\ref{fig:J_Comp}. First, we perform a binned Jeans analysis (see
  \citealt{2015arXiv150402048B}) of the kinematic data, and find
  compatible results. Second, we calculate the bootstrap mean and
  dispersion of the $J$-factor \citep{1982jbor.book.....E}. For this
  purpose, we generate 500 bootstrap resamples\footnote{The best-fit
    DM profile and anisotropy parameters for each sample are obtained
    by maximizing the likelihood of Eq.~(\ref{eq:likelihood}). J-factors were then computed for these best fitting profiles.} by drawing with replacement 16 stars among the 16 of the original sample with $P_i > 0.95$. The results are in excellent agreement with the MCMC analysis. Finally, we use all 38 stars of the sample but weight the log-likelihood function of Eq.~(\ref{eq:likelihood}) by the membership probabilities $P_i$ \citep{2015arXiv150402048B}. As only one star shows an intermediate membership probability $0.1 < P_i < 0.95$, we obtain very similar results. These two tests confirm that the reconstruction of the astrophysical factors of Ret~II is not significantly affected by outliers. This is not always the case, notably for Segue~I (Bonnivard, Maurin \& Walker, in prep.).

We note that \citet{2015arXiv150402889S} independently performed
  an analysis of the M2FS Ret~II spectroscopic data and found a slightly smaller $J$-factor. 
This can be traced to their choice of priors and light profile (L. Strigari, private communication). A detailed comparison will be presented in Geringer-Sameth at al. (in prep.).
\section{Comparison to other dSphs}
\label{sec:Comp}
The same Jeans analysis has been applied to twenty-one other dSphs in
\citet{2015arXiv150402048B}. In Figure \ref{fig:J_Comp}, we compare
the $J$-factors (for $\alpha_{\rm int} = 0.5^{\circ}$) of Ret~II to
the brightest objects identified in
\citet{2015arXiv150402048B}\footnote{Segue~I may have a highly
  uncertain $J$-factor (Bonnivard, Maurin \& Walker, in prep.). We
  show it only for illustration purposes.}. Ret~II is comparable to
Wilman~I in terms of its median $J$-factor, but slightly below Coma Berenices
and Ursa Major~II. Its CIs are typical of an `ultrafaint' dSph, and significantly larger than the uncertainties of `classical' dSphs.

Interpreting the possible $\gamma$-ray signal in Ret~II in terms of DM annihilation \citep{2015arXiv150302320G,2015arXiv150306209H}, one would expect similar emissions from the dSphs with comparable $J$-factors, such as UMa~II, Coma, and Wil~I. However, no excess was reported from these latter objects \citep{2014arXiv1410.2242G,2015arXiv150302641F}. This could be explained by the large statistical and systematic\footnote{The latter comes from a possible triaxiality of the dSph (0.4 and 0.3 dex for annihilation and decay respectively, see \citealt{2015MNRAS.446.3002B}), and depends on the l.o.s. orientation with respect to the principle axes of the halo.} uncertainties in the J-factors. Moreover, the Jeans analysis assumes all of these objects to be in dynamical equilibrium, but tidal interactions with the Milky Way could artificially inflate the velocity dispersion and therefore the astrophysical factors. UMa~II, and to a lesser extent Coma, appear to be experiencing tidal disturbance \citep{2007ApJ...670..313S,2007MNRAS.375.1171F,2010AJ....140..138M,2013MNRAS.433.2529S}, while Wil~I may show non-equilibrium kinematics \citep{2011AJ....142..128W}. Caution is therefore always advised when interpreting the astrophysical factors of these objects. The dynamical status of Ret~II is not yet clear.  Its flattened morphology may signal ongoing tidal disruption.  However, the available kinematic data do not exhibit a significant velocity gradient that might be associated with tidal streaming motions \citep{kinematics}.

\section{Conclusion}
We have applied a spherical Jeans analysis to the newly discovered
dSph Ret~II, using sixteen likely members from the kinematic data set
of \citet{kinematics}. We employed the optimized setup of
\citet{2015MNRAS.446.3002B,2015arXiv150402048B}, which was found to
mitigate several biases of the analysis, and checked that our results
are robust against several of its ingredients. We find that Ret~II
presents one of the largest annihilation $J$-factors among the Milky
Way's dSphs, possibly making it one of the best targets to constrain
DM particle properties. However, it is important to obtain follow-up
photometric and spectroscopic data in order to test the assumptions of
dynamical equilibrium as well as to constrain the fraction of binary
stars in the kinematic sample. Nevertheless, the proximity of Ret II
and its apparently large dark matter content place it among the most
attractive targets for dark matter particle searches.

\acknowledgments This work has been supported by the ``Investissements d'avenir, Labex
ENIGMASS", and by the French ANR, Project DMAstro-LHC,
ANR-12-BS05-0006. MGW is supported by National Science Foundation grants AST-1313045, AST-1412999. SMK is supported by DOE DE-SC0010010, NSF PHYS-1417505, and NASA NNX13AO94G. MM is
supported by NSF grants AST-0808043 and AST-1312997. EWO is supported by NSF grant AST-0807498 and AST-1313006.


\end{document}